\documentclass[a4paper,11pt]{article}
\usepackage{pos}

\title{Emergent phenomena from centre vortices}

\author*[a]{Waseem Kamleh}
\author[a]{James Biddle}
\author[a]{Derek B. Leinweber}
\author[a,b]{Finn M. Stokes}

\affiliation[a]{Centre for the Subatomic Structure of Matter, Department of Physics,\\ The University of Adelaide, SA 5005, Australia}
\affiliation[b]{J\"ulich Supercomputing Centre, Institute for Advanced Simulation,\\Forschungszentrum J\"ulich, J\"ulich D-52425, Germany}

\emailAdd{waseem.kamleh@adelaide.edu.au}
\emailAdd{james.biddle@adelaide.edu.au}
\emailAdd{derek.leinweber@adelaide.edu.au}
\emailAdd{f.stokes@fz-juelich.de}

\abstract{%
Quark confinement is perhaps the most important emergent property of the theory of quantum chromodynamics. Herein we review some key aspects of centre vortices in SU(3) lattice gauge theory. Starting from the original Monte Carlo gauge fields, a vortex identification
procedure yields vortex-removed and vortex-only backgrounds. The comparison between the original `untouched' Monte Carlo gauge fields and these so called vortex-modified ensembles has provided a variety of results that support the notion that centre vortices are fundamental to confinement in pure gauge theory. For the first time we perform direct numerical tests of the response of centre vortices to the presence of dynamical quarks in SU(3).}

\FullConference{%
 The 38th International Symposium on Lattice Field Theory, LATTICE2021
  26th-30th July, 2021
  Zoom/Gather@Massachusetts Institute of Technology
}

\DeclareMathOperator{\re}{Re}
\DeclareMathOperator{\Tr}{Tr}

\newcommand{\braket}[1]{\langle #1 \rangle}

\usepackage{tikz,varwidth}
\usepackage{tikzscale}
\usetikzlibrary{shapes.geometric, patterns}
\usetikzlibrary{arrows.meta,shapes,automata,petri,positioning,calc,decorations.markings}

\begin{document}

\maketitle

\section{Introduction}

What defines emergent phenomena? Put simply, an emergent behaviour or emergent property can appear when a number of simple entities (agents) operate in an environment, forming more complex behaviours as a collective. Quark confinement and dynamical chiral symmetry breaking appear to be emergent properties of QCD, generally accepted to originate from topological structures present in the nontrivial vacuum. An analytic proof deriving the underlying mechanisms responsible for these phenomena from the QCD Lagrangian has yet to be achieved. Nonetheless, significant progress has been made towards identifying the origin of quark confinement through lattice field theory, which is well-suited to the kind of non-pertubative simulations required to study quantum chromodynamics.

The precise form of the topological objects that result in confinement remains a topic of discussion. Here we focus on centre vortices. There is now a significant body of evidence from lattice simulations that centre vortices are the prime candidate for the mechanism underpinning quark confinement. The centre vortex model of confinement~\cite{'tHooft:1977hy,'tHooft:1979uj,DelDebbio:1996lih,Faber:1997rp,DelDebbio:1998luz,Bertle:1999tw,Faber:1999gu,Engelhardt:1999xw,Bertle:2000qv,Greensite:2003bk,Engelhardt:2003wm,Greensite:2016pfc,Cornwall:1979hz,Nielsen:1979xu,Ambjorn:1980ms,Vinciarelli:1978kp,Yoneya:1978dt,Mack:1978rq} was proposed many years ago now, and is well-known.

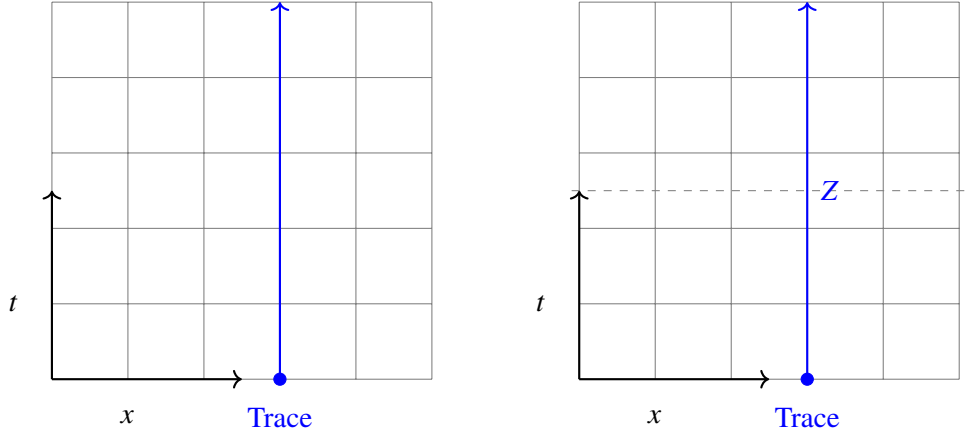
\begin{figure}
\centering
\null\hfill
\begin{tikzpicture}
  \draw [step=1.0,blue, thin, gray] (0,0) grid (5,5);
  \draw [thick, blue, ->] (3,0) -- (3,5);
  \draw [thick, ->] (0,0) -- (2.5,0);
  \draw [thick, ->] (0,0) -- (0,2.5);
  \draw [blue, fill] (3,0) circle (0.8mm);
  \node at (-0.5,1.0) {$t$};
  \node at (1.0,-0.5) {$x$};
  \node [blue] at (3.0,-0.5) {Trace};
\end{tikzpicture}
\hfill
\begin{tikzpicture}
  \draw [step=1.0,blue, thin, gray] (0,0) grid (5,5);
  \draw [thick, blue, ->] (3,0) -- (3,5);
  \draw [thick, ->] (0,0) -- (2.5,0);
  \draw [thick, ->] (0,0) -- (0,2.5);
  \draw [blue, fill] (3,0) circle (0.8mm);
  \node at (-0.5,1.0) {$t$};
  \node at (1.0,-0.5) {$x$};
  \node [blue] at (3.0,-0.5) {Trace};
  \draw [thin, gray, dashed] (-0.1,2.5) -- (5.1,2.5);
  \node [blue] at (3.3,2.5) {$Z$};
\end{tikzpicture}
\hfill\null
\caption{Representation of a Polyakov loop (left), and the same loop under a global centre transformation $Z = zI$ (right). The Polyakov loop is closed via the periodic boundary conditions and hence gauge invariant. Under a global centre transformation the Polyakov loop acquires a phase equal to the centre phase $z.$}
\label{fig:polyakov}
\end{figure}
\begin{figure}[t]
  \centering
  \includegraphics[trim=35mm 0mm 35mm 0mm,clip,width=0.48\textwidth]{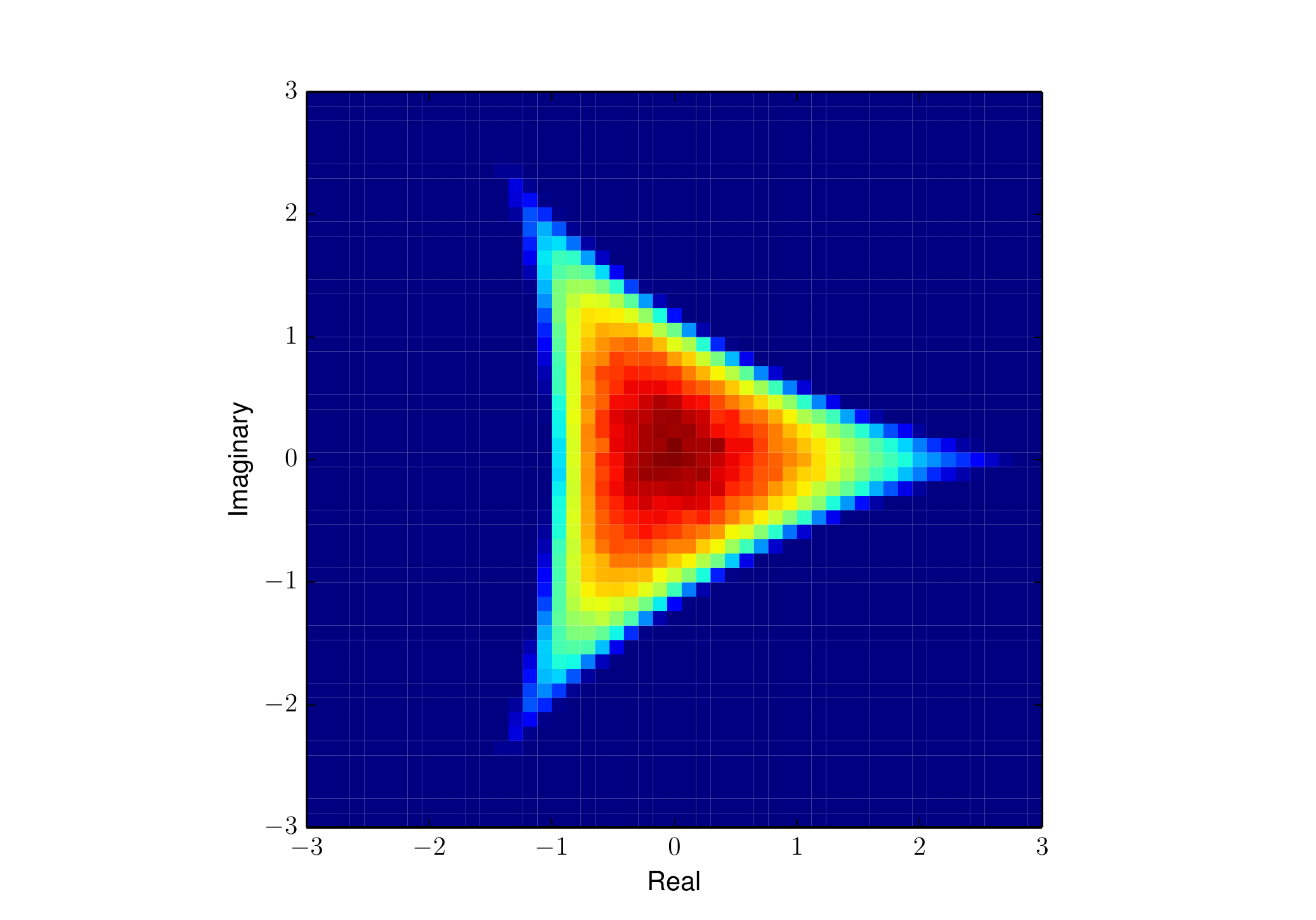}
  \includegraphics[trim=35mm 0mm 35mm 0mm,clip,width=0.48\textwidth]{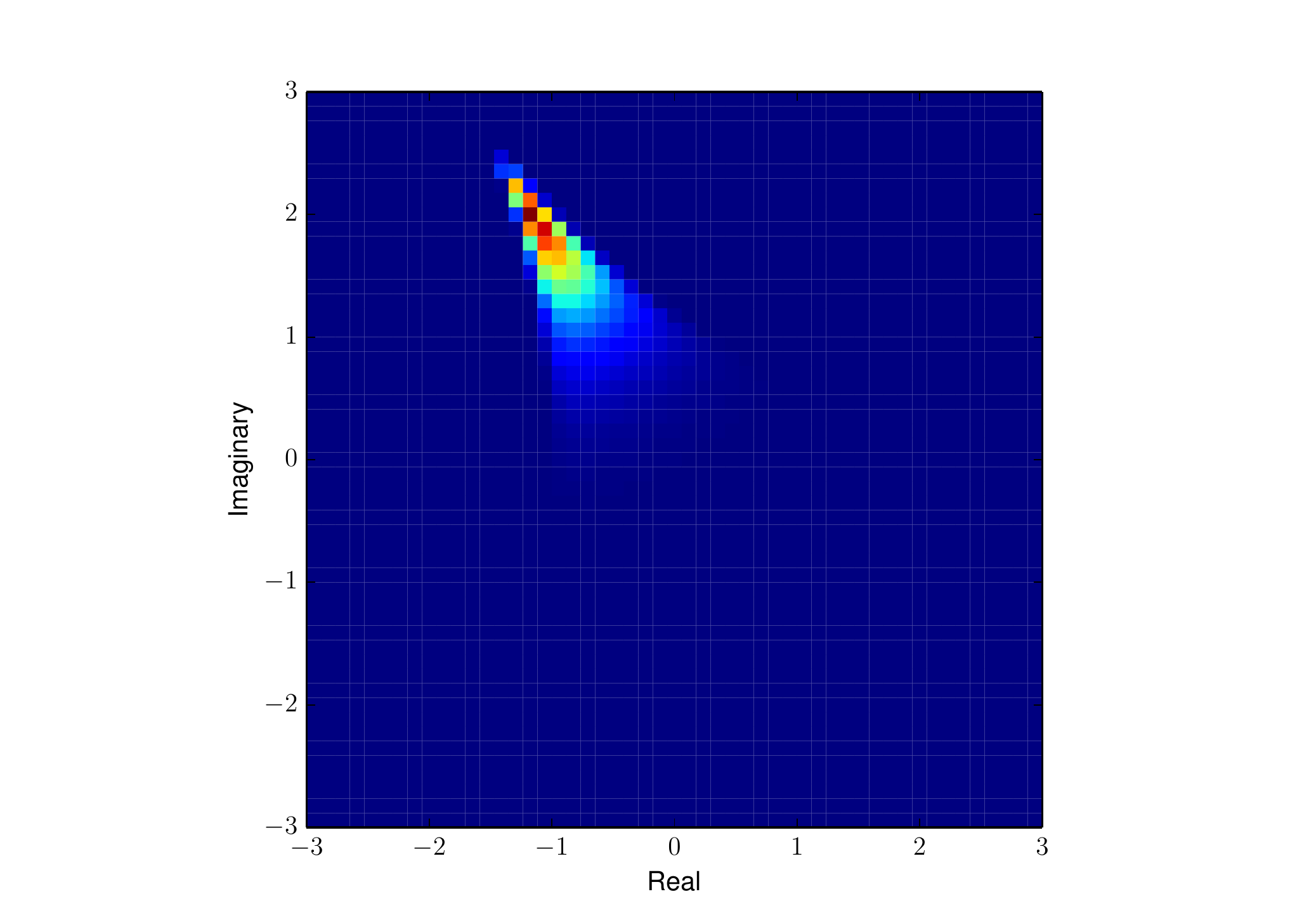}
  \caption{Distribution of Polyakov loop \(L(\vec{x})\) values on a pure gauge lattice configuration, in the confining phase below the critical temperature \(T_C\) (left); and in the deconfining phase above \(T_C\) (right).}
  \label{fig:polyakovphase}
\end{figure}

%
The significance of centre vortices and their relation to confinement in $SU(N)$ gauge theory is most easily understood on the lattice in the context of the Polyakov loop (see Fig.~\ref{fig:polyakov}), which is the trace of the product of $N_t$ temporal links across the lattice (and hence gauge invariant by the periodic boundary),
\begin{equation}
L(\vec{x}) = \Tr \prod_t U_0(t,\vec{x}). %
\end{equation}
The Polyakov loop can be thought of as the world-line of a massive static quark, i.e. $\langle L(\vec{x}) \rangle = \exp(-F_{q}N_{t})$ where $F_q$ is the quark free energy.
\begin{itemize}
\item In a confining phase, the free energy $F_{q} \rightarrow \infty,$ and $\langle L(\vec{x}) \rangle=0.$
\item In a deconfining phase, the free energy $F_{q}$ is finite, and $\langle L(\vec{x}) \rangle$ is non-zero.
\end{itemize}
Hence, the Polyakov loop is $\braket{L(\vec{x})}$ is an order parameter for confinement. Now, consider a global centre transformation on some timeslice,
\begin{equation}
  U_{0}(t_0,\vec{x}) \rightarrow Z\,U_{0}(t_0,\vec{x}), \quad \forall \vec{x}.
  \label{eq:globalz}
\end{equation}
\begin{figure}
\centering
\null\hfill
\begin{tikzpicture}
  \draw [step=1.0,blue, thin, gray] (0,0) grid (5,5);
  \draw [thick, blue, ->] (2,2) -- (3,2) -- (3,3) -- (2,3) -- (2,2);
  \draw [blue,fill] (2,2) circle (0.8mm);
  \node [blue] at (2.0,1.7) {Trace};
  \draw [thick, ->] (0,0) -- (1.5,0);
  \draw [thick, ->] (0,0) -- (0,1.5);
  \node at (-0.5,1.0) {$t$};
  \node at (1.0,-0.5) {$x$};
\end{tikzpicture}
\hfill
\begin{tikzpicture}
  \draw [step=1.0,blue, thin, gray] (0,0) grid (5,5);
  \draw [thin, gray, dashed] (-0.1,2.5) -- (5.1,2.5);
  \draw [thick, blue, ->] (2,2) -- (3,2) -- (3,3) -- (2,3) -- (2,2);
  \draw [blue,fill] (2,2) circle (0.8mm);
  \node [blue] at (2.0,1.7) {Trace};
  \node [blue] at (1.7,2.5) {$Z^*$};
  \node [blue] at (3.3,2.5) {$Z$};
  \draw [thick, ->] (0,0) -- (1.5,0);
  \draw [thick, ->] (0,0) -- (0,1.5);
  \node at (-0.5,1.0) {$t$};
  \node at (1.0,-0.5) {$x$};
\end{tikzpicture}
\hfill\null
\caption{Representation of a plaquette (left), and the same plaquette under a global centre transformation $Z = zI$ (right). The plaquette is invariant under the global centre transformation as the centre elements $Z$ and $Z^*$ commute with the gauge links, and $Z^* Z = I.$ }
\label{fig:plaquette}
\end{figure}
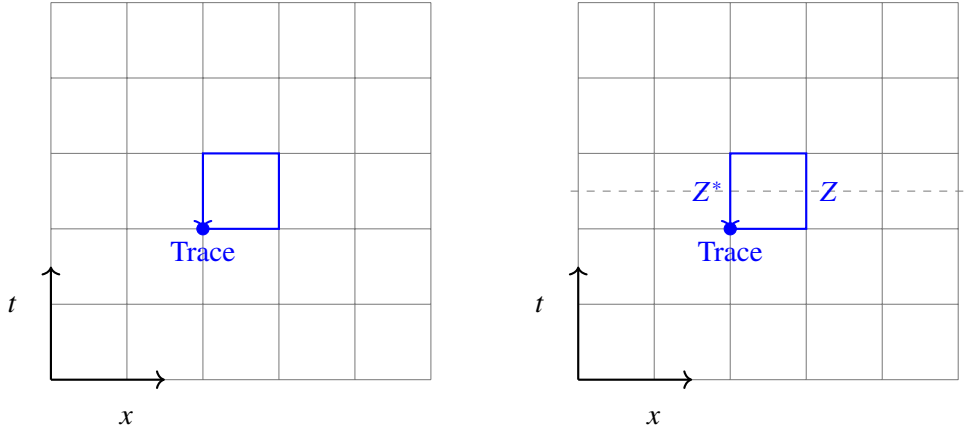%
In a pure $SU(N)$ gauge theory, the Yang-Mills action is proportional to the sum over the lattice of the real part of the trace of the unique plaquettes (here we choose to set $N=3$ for clarity),
\begin{equation}
  S_{YM} = \beta\sum_{x,\mu < \nu}\left(1-\frac{1}{3}\re\Tr\left[ U_\mu(x)U_\nu(x+\hat{\mu})U_\mu^\dag(x+\hat{\nu})U_\nu^\dag(x) \right]\right).
\end{equation}
As demonstrated in Fig.~\ref{fig:plaquette}, the global centre transformation above is a symmetry of the Yang-Mills gauge action $S_{YM}$ as long as $Z$ commutes with $U_{\mu}(x)$, i.e. $Z$  is in the centre group of $SU(3),$
\begin{equation}
  Z = \exp(m\,2\pi i/3) \equiv z\mathrm{I},\quad m \in \{-1,0,1\}.
\end{equation}
Under this centre transformation, as shown in Fig.~\ref{fig:polyakov} the Polyakov loop acquires a phase,
\begin{equation}
L(\vec{x}) \rightarrow z L(\vec{x}).
\end{equation}
It is straightforward from the above to see that in the confining phase where the expectation value of the Polyakov loop vanishes, that expectation value is invariant under a centre transformation. Conversely, in the deconfining phase where the expectation value of the Polyakov loop is finite, then that value is not preserved by a centre transformation. These two situations are illustrated in Fig.~\ref{fig:polyakovphase}, which shows a heatmap of the distribution of the Polyakov loop values in the two different phases. In the left plot showing the confining phase, the centre symmetry of the Polyakov loop is clear. In the right plot in the deconfining phase, there is a preferred centre phase and hence no centre symmetry. It has been stated elsewhere that \emph{``confinement is the phase of unbroken global center symmetry''}~\cite{Greensite:2016pfc}, and this statement can certainly be considered reasonable in the context of a pure $SU(N)$ gauge theory.

\section{Centre vortex identification}

Through the study of so-called \emph{vortex-modified} ensembles in $SU(3)$ lattice gauge theory, there is now an extensive body of evidence supporting the idea that centre vortices are responsible for quark confinement and dynamical chiral symmetry breaking. On the lattice we seek to decompose the gauge links $U_\mu(x)$ in the following form
\begin{equation}
U_{\mu}(x) = Z_{\mu}(x)\cdot R_{\mu}(x),
\end{equation}
such that the vortex content is captured in the field of centre-projected elements $Z_{\mu}(x).$ The remaining short-range fluctuations are described by the vortex-removed field $R_{\mu}(x).$ By fixing to Maximal Centre Gauge~\cite{DelDebbio:1996lih,Langfeld:1997jx,Langfeld:2003ev,Cais:2008za} and then projecting the gauge-fixed links to the nearest centre element, we may identify the vortex matter by searching for plaquettes with a nontrivial centre flux around the boundary. These are the ``thin'' centre vortices (or P-vortices), that are embedded within the physical ``thick'' centre vortices of the original, Monte Carlo generated, configurations. The centre-projection procedure creates three distinct ensembles of SU(3) gauge fields:
\begin{enumerate}
\item The original `untouched' configurations, $U_{\mu}(x),$

\item The projected vortex-only configurations, $Z_{\mu}(x),$

\item The vortex-removed configurations, $R_\mu(x) = Z^{\dagger}_{\mu}(x)\,U_{\mu}(x).$
\end{enumerate}
Performing lattice calculations of key quantities on the three different ensembles and comparing the differences that emerge from the absence or presence of centre vortices has proven to be a rich source of information regarding the properties of the $SU(3)$ Yang-Mills theory, which we also refer to as the \emph{pure gauge} theory when it needs to be distinguished from the theory containing dynamical fermions.

\section{Centre vortices and dynamical fermions}

Lattice simulations on vortex-modified ensembles in pure gauge theory have established a significant body of evidence supporting the argument that centre vortices are the fundamental mechanism underpinning the confinement of quarks and dynamical chiral symmetry breaking. These studies have examined a number of different areas:
\begin{itemize}
\item Static quark potential investigations~\cite{Langfeld:2003ev,Cais:2008za, Trewartha:2015ida} have revealed that vortex removal removes the linear potential. In $SU(3)$ the vortex-only field recreates $\sim 2/3$ of the untouched string tension. 
\item A connection to instanton degrees of freedom has been established via smoothing~\cite{Trewartha:2015ida,Biddle:2019gke}. Vortex removal is shown to destabilise instantons under cooling, whereas the vortex-only field creates instantons under cooling. This leads to the notion that centre vortices contain the 'seeds' of instantons.
  \item The overlap quark propagator possesses good chiral properties, allowing for studies of the quark mass function on vortex-modified ensembles~\cite{Trewartha:2015nna}. These show that vortex removal causes loss of dynamical mass generation, and the vortex-only field (with some smoothing) reproduces dynamical mass generation.
\item Examining the hadron spectrum on vortex-modified ensembles~\cite{Trewartha:2017ive,Trewartha:2015nna,OMalley:2011aa} shows that vortex removal restores chiral symmetry at light masses, signalled by the emergent degeneracy of hadronic channels related by a chiral transformation. Furthermore, it is shown that the vortex-only field can reproduce the structure of the ground state hadron spectrum.
\item Recent studies of the gluon propagator~\cite{Biddle:2018dtc} have shown that vortex removal causes a loss of infrared strength. The vortex-only field recreates $\sim 2/3$ of the infrared strength of the untouched gluon propagator. 
\end{itemize}
It is remarkable that the key emergent features of QCD remain after reducing the eight real parameters of an $SU(3)$ matrix to the 3 discrete values that characterise a vortex in $Z(3).$ Indeed, 't Hooft stated~\cite{'tHooft:1977hy} that \emph{``it would be tempting to abolish the $SU(3)$ color theory for hadrons altogether,
 replacing it by a $Z(3)$ theory on a Euclidean lattice and taking the continuum limit close to the critical point.''} Given the evidence that has been accumulated by lattice studies in pure gauge theory, this statement seems remarkably prescient. 

However, the response of vortices to the presence of dynamical fermions is fertile ground for exploration. There have been a number of investigations within the context of gauge-Higgs theory~\cite{Fradkin:1978dv,Bertle:2003pj,Greensite:2006ng,Greensite:2017ajx,Greensite:2018mhh,Greensite:2020nhg}. Here and in other reports within these conference proceedings we present the first numerical tests of centre vortices in an $SU(3)$ gauge theory with dynamical fermions. Our results are generated using the PACS-CS $(2+1)$-flavour full-QCD ensembles~\cite{PACS-CS:2008bkb} made available through the ILDG~\cite{Beckett:2009cb}. These ensembles use a $32^3 \times 64$ lattice volume, with a renormalisation-group improved Iwasaki gauge action and non-perturbatively improved clover fermion action. For comparison, we have created a matched $32^3 \times 64$ pure-gauge ensemble using the same improved Iwasaki gauge action at a similar lattice spacing.

\begin{figure}
  \includegraphics[width=0.9\textwidth]{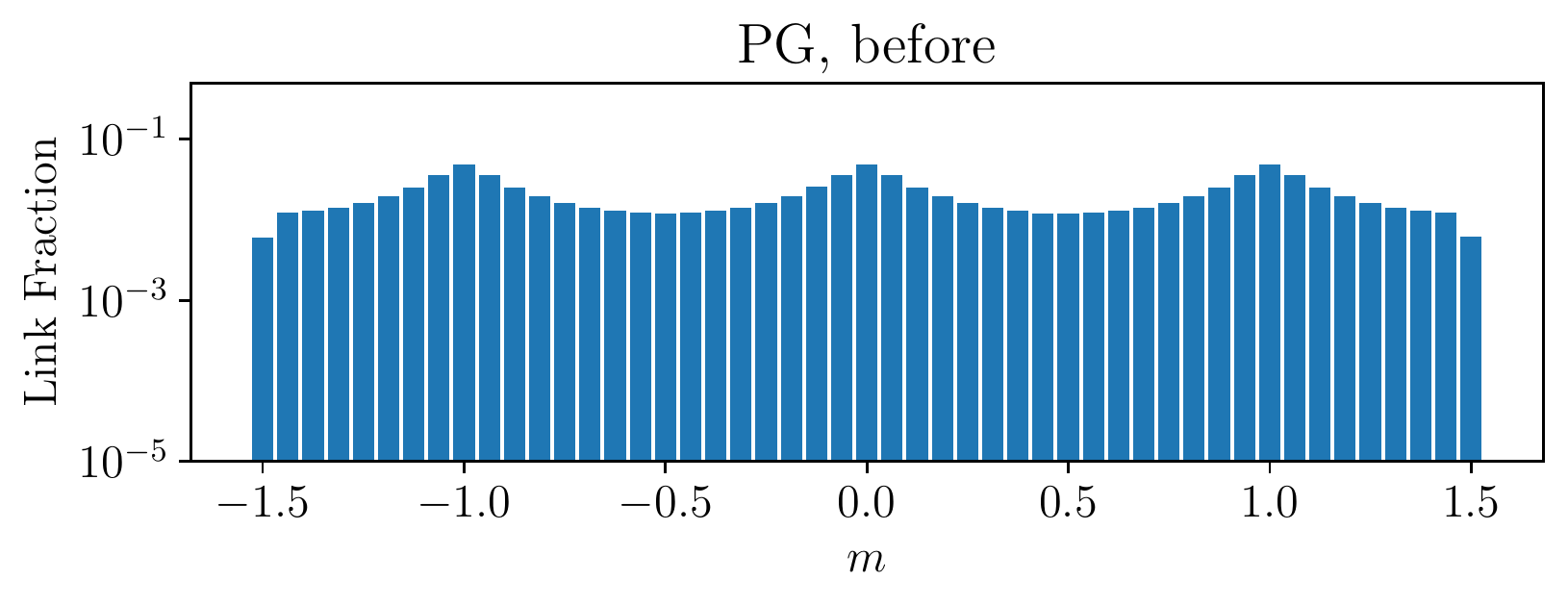}\\
  \includegraphics[width=0.9\textwidth]{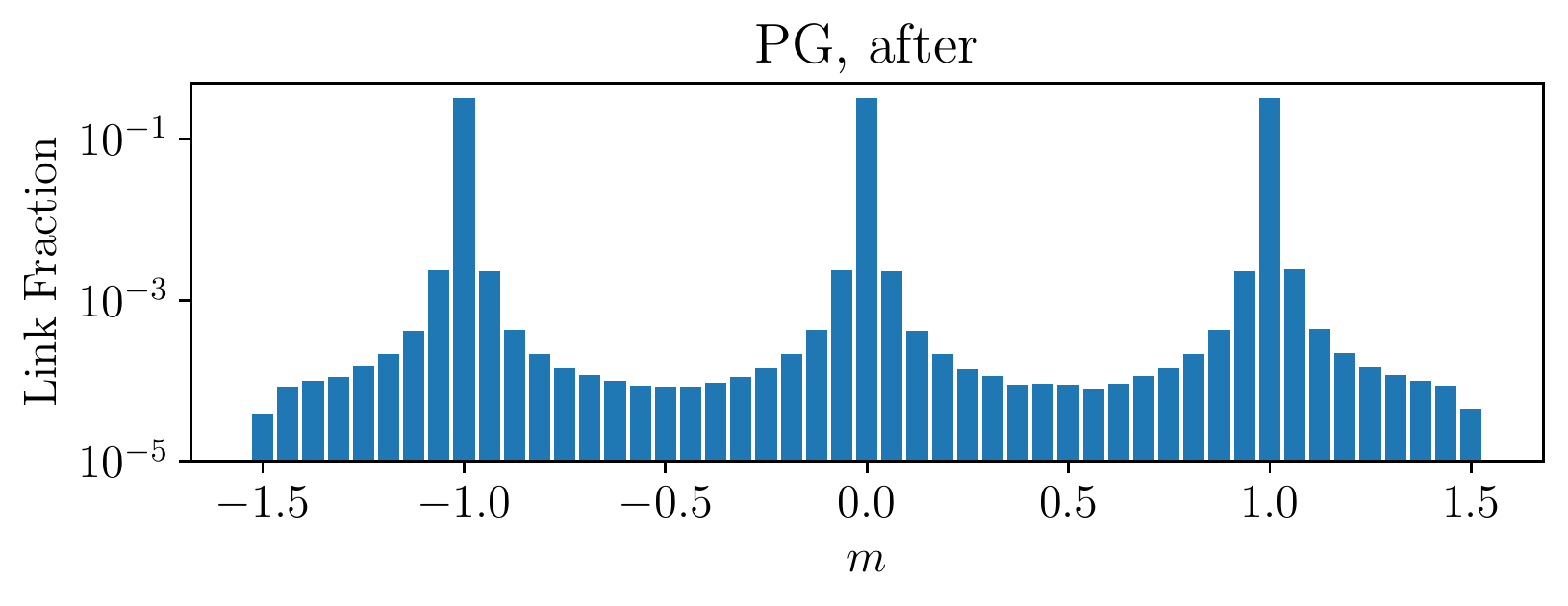}
  \caption{Distribution of the phase of the trace of the links on a single pure gauge field configuration, before (top) and after (bottom) MCG projection.}
  \label{fig:PGphase}
\end{figure}
Figure~\ref{fig:PGphase} shows the distribution of the phase of the trace of the links on a single pure gauge configuration, $\Tr U = r \exp (m\,2\pi i/3),$ before and after Maximal Centre Gauge (MCG) projection. As would be expected within the pure gauge theory which possesses a global centre symmetry, we see that there is an equal distribution peaked around the three centre phases $m = -1, 0, +1.$ Note the logarithmic scale on the vertical axis, which indicates that after MCG projection the phase distribution is sharply peaked around the three centre phases.

\begin{figure}
  \includegraphics[width=0.9\textwidth]{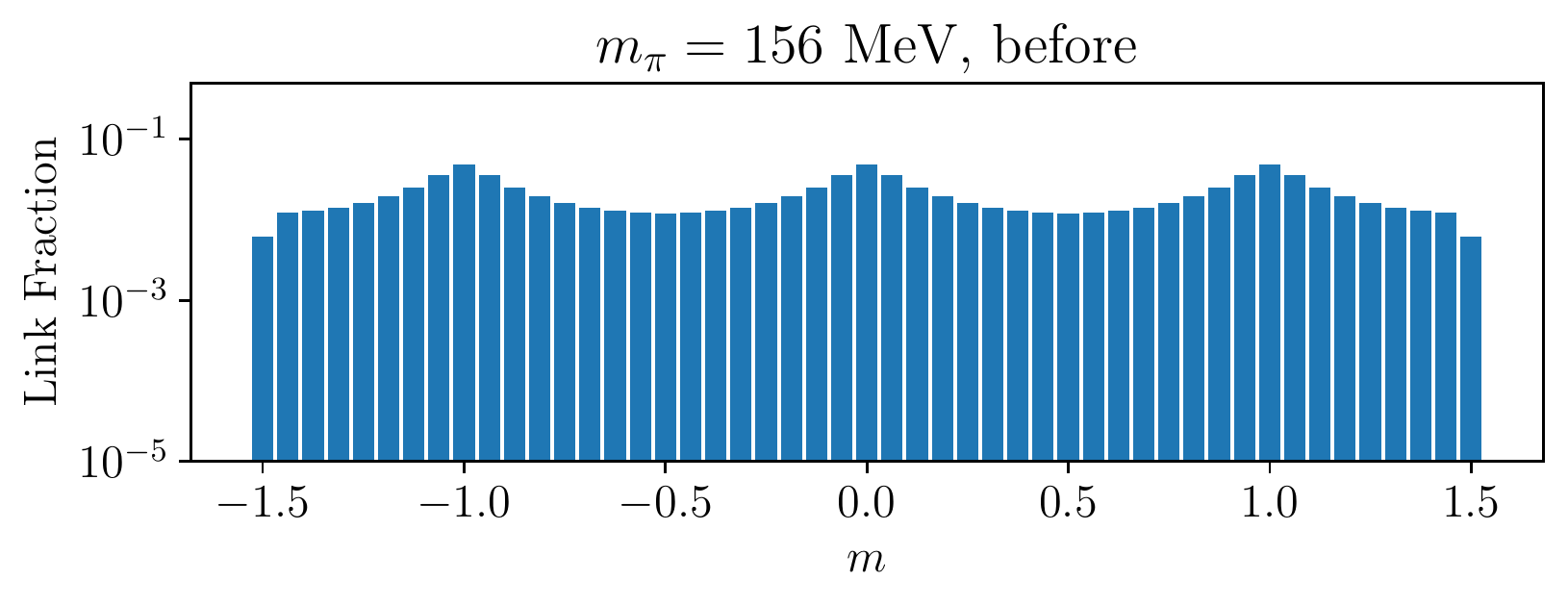}\\
  \includegraphics[width=0.9\textwidth]{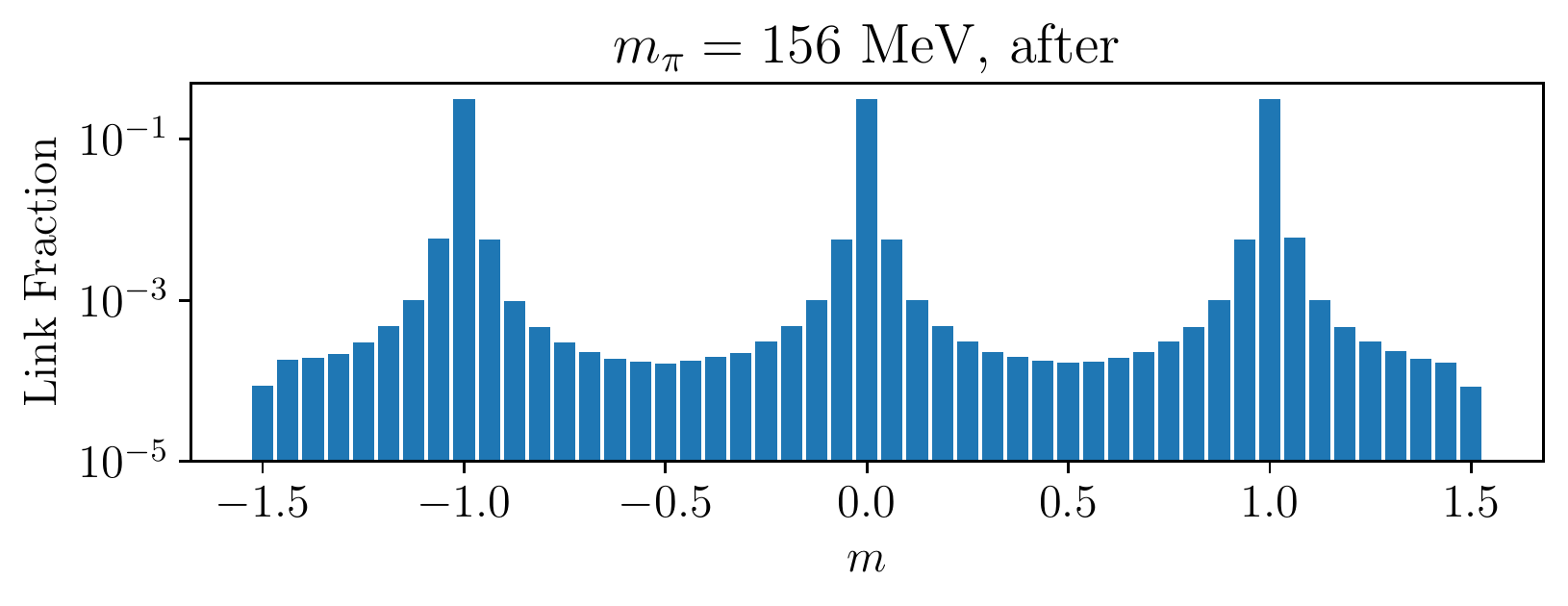}
  \caption{Distribution of the phase of the trace of the links on a single dynamical gauge field configuration, before (top) and after (bottom) MCG projection.}
  \label{fig:DYNphase}
\end{figure}
Figure~\ref{fig:DYNphase} shows the distribution of the phase of the trace of the links on a single dynamical fermion configuration at the lightest pion mass of $m_\pi = 156$ MeV, before and after Maximal Centre Gauge (MCG) projection. It is indeed remarkable that again, even in the presence of matter fields there is an equal distribution peaked around the three centre phases $m = -1, 0, +1.$ This is perhaps somewhat surprising, given that the global centre symmetry that was defined above for the pure gauge theory is no longer a symmetry when the quark fields are introduced~\cite{'tHooft:1977hy}.

The notion that quark confinement (or not) is the result of competing order and disorder effects was introduced early on~\cite{'tHooft:1977hy}, and since then there has been significant efforts to refine what precisely is meant by the term ``confinement'' both with and without the presence of matter fields (see for example~\cite{Greensite:2003bk,Greensite:2017ajx,Greensite:2018mhh,Greensite:2020nhg}). In a refinement of the statement quoted earlier at the end of the first section, what we can perhaps infer from these results is that \emph{confinement is an emergent feature of QCD that is realised when the vacuum phase embodies global centre symmetry.} Nonetheless, a better understanding will only come from further investigation to be undertaken in future studies.

\section*{Acknowledgements}

We thank the PACS-CS Collaboration for making their 2 +1 flavour configurations available via the International Lattice Data Grid (ILDG).
This research was undertaken with the assistance of resources from the National Computational Infrastructure (NCI), provided through the
National Computational Merit Allocation Scheme and supported by the Australian Government through Grant No. LE190100021 via the
University of Adelaide Partner Share. This research is supported by Australian Research Council through Grants No. DP190102215 and DP210103706.
WK is supported by the Pawsey Supercomputing Centre through the Pawsey Centre for Extreme Scale Readiness (PaCER) program.


\providecommand{\href}[2]{#2}\begingroup\raggedright\endgroup

\end{document}